\documentstyle[a4,12pt]{article}
\input{psfig.sty}

\begin{document}

\title{Two--Component Dust in  Spherically Symmetric Motion}
\author{Gernot Haager \\
 Institute of Theoretical Physics  \\
 Friedrich--Schiller--University \\
 Fr{\"o}belstieg 1, 07743 Jena, Germany}
\maketitle
\abstract{Two  components of spherically symmetric,
inhomogeneous
dust penetrating  each other
 are introduced as a generalization of the well--known Tolman--Bondi
dust solution. 
The field equations of this model are 
formulated and general properties
are discussed. 
  Special solutions with additional symmetries
 --- an extra Killing-- or homothetic vector ---
 and their matching to the corresponding Tolman--Bondi solution are
 investigated.
 } \\
 PACS Numbers: 04.20.Jb, 04.40.Nr
\section{Introduction}
 The Tolman--Bondi solution (TBS) 
 for spherically symmetric has been found by Lemaitre \cite{Lem1}
 1933 and Tolman \cite{Tol1} 1934
 as one of the first inhomogeneous
 cosmological models. Its physical properties has been widely
 investigated, e.g. Bondi 1947 \cite{Bondi1}.
It has been used as a model to study the
 gravitational collapse and the formation of a black hole.
 In recent years spherically symmetric dust was studied to
 simulate such phenomena like shell--crossing \cite{Mesz1} or voids \cite{Bon1}.
 The TBS has  also been taken as background metric for thin shells of dust 
 to model  voids and thin shells \cite{Maeda1}.

Spherically symmetric dust has been generalized in various directions.
Mixtures of dust and radiation \cite{Per1} or dust and a perfect fluid
\cite{Suss1} 
have
been investigated.
Coley and Tupper \cite {Col1} introduced a two--fluid model for general
viscous fluids and examined various special cases.
In this paper we consider  dust shells of
finite size moving with different four velocities. So two shells of dust can move
towards each other to form a common region where two dust
components exist simultaneously. To model voids spherically symmetric dust with
different values for the energy densities has been matched by an intermediate
region. Here comoving boundary surfaces have been considered. By introducing two dust
components the boundaries need no longer to be comoving.
Furthermore one can examine 
a central core of two dust components with a mass flux through
 the boundary that
can be matched
to an exterior one--component dust.

A solution for the two--component dust (TCD)  has to solve
a coupled nonlinear system of partial differential equations (PDEs)
in two independent variables for two functions.  To find an explicit
solutions
 further assumptions as
additional symmetries   are made. 
These
special solutions still  allow us to study different physically and 
geometrically interesting
space times and  lead to different topologies.

The paper is organized as follows. 
In section 2  the Tolman--Bondi solution is given.
The model of two--component dust  is introduced in section 3.
The field equations are formulated in an invariant way and
general properties of the metric 
 are discussed. Section four deals with
the matching procedure.
In the sections 5 and 6 some special solution with an extra Killing--
or homothetic vector in the two--component dust region are examined in detail. 

\section{One component of spherically symmetric dust}
\setcounter{equation}{0}
For dust, i.e. a perfect fluid 
without pressure, the Einstein field equations read
\[ R_{ab} - \frac{1}{2} R \, g_{ab} = T_{ab}, \]
with the energy momentum tensor 
\begin{equation}
T_{ab} = \rho \, u_a u_b , \quad u_a u^a = -1 , \quad u_{a;b} u^b = 0,
\label{Tab1}
\end{equation}
where $\rho$ is the energy density and $u_a$ a time-like geodesic vector
field.
In a comoving spherically symmetric coordinate system the
metric 
can be written in the form
\begin{equation}
ds^2 = Y^2 (r,t) d\Omega^2 + R^2(r,t) dr^2 - dt^2
\label{tolman1}
\end{equation}
where $d\Omega^2$ is the usual line element on the unit sphere.
Here the solution of the field equations can be given explicitly, see e.g. \cite{Kra1}.

For a non--comoving frame of reference, which we will use later 
for the matching procedure, the line
element is best given by 
\begin{equation}
ds^2 = Y^2 d\Omega^2 + \frac{1}{\dot{Y}^2 + Y_{,a} Y^{,a} } \left[
       (Y_{,a} + \dot{Y} u_a) dx^a (Y_{,b} + \dot{Y} u_b) dx^b \right]
	      - u_a u_b  dx^a dx^b.
\label{tolmaninv1}
\end{equation}
Here 
$ Y_{,a}$ is the gradient of the sphere radius $Y$
and
\[ \dot{Y} = Y_{,a} u^a  = {\cal L}_u Y \]
the derivative of  $Y$ along $u_a$.
The free functions $M$ and $f$ can be expressed by
\begin{eqnarray}
M & = & Y ( 1- Y_{,a} Y^{,a} ),
\label{mass1} \\
f & = & -1 + \dot{Y}^2 + Y_{,a} Y^{,a}.
\label{para1}
\end{eqnarray}
Their specific coordinate dependence is defined
 by
\begin{equation}
{\cal L}_u M = \dot{M} = 0, \quad {\cal L}_u f = \dot{f} = 0,
\label{paradep1}
\end{equation}
as $f$ and $M$ are constant along the fluid's world lines.
With the expressions (\ref{mass1}), (\ref{para1}) and the metric
(\ref{tolmaninv1})
the field equations   for one--component 
dust are fulfilled identically.

The sign of the free function $f$, sometimes called the "energy function",
leads to three different types of solutions. The hyperbolic case,
$f> 0$, represents an open unbound model in contrast to the
elliptic case, $-1 < f <0$, which is a bound one. The parabolic case,
$f=0$, is marginally bound.
The  function $M$ gives twice the mass inside a
sphere with radius $Y$ 
so it is usually  called the "mass function".

\section{The model}
\setcounter{equation}{0}
The energy momentum tensor for two--component  dust in radial motion is
\begin{equation}
T_{ab} = \mu \, v_a v_b + \epsilon \, u_a u_b .
\label{Tab2}
\end{equation}
We assume that both dust components are only coupled by their gravitational
attraction  and the mass density is conserved for each component, so
they have to move on geodesics:
\begin{equation}
v_a v^a = -1 \quad v_{a;b} v^b = 0 \quad \mbox{and} \quad 
u_a u^a = -1 \quad u_{a;b} u^b = 0.
\end{equation}

For this model one can again introduce a mass function $M$ given by (\ref{mass1})
which was first considered
 by Lemaitre \cite{Lem1} and rediscovered by Misner
\cite{Mis1} for perfect fluids. $M$ represents the active gravitational mass
and two field equations can be reformulated with the aid of $M$.
The rate of change for the mass function along each fluid line
is given by
\begin{eqnarray}
M_{,a} v^a & = & \epsilon Y^2 Y_{,a} \left( v^a + u_b v^b u^a\right)
\label{Mfeld1} \\
M_{,a} u^a & = & \mu Y^2 Y_{,a} \left( u^a + u_b v^b v^a \right)
\label{Mfeld2}. % mistake in original version
\end{eqnarray}
$M$ is only conserved for one component if the other component vanishes.
The remaining field equations not containing the energy densities
can be formulated in an invariant way, too. The first one requires
that the
tangential pressure  vanishes, $G_{\theta \theta}=0$, and
the second one  reads
\begin{equation}
-2 Y Y_{;a;b} u^a v^b + \frac{M}{Y} u_a v^a = 0 
\label{feldgl1inv1}
\end{equation}
which is a consequence that $u^a$ and $v^a$ are both geodesic.

The line element can be  written as
\begin{equation}
ds^2 = Y^2 d\Omega^2 + \frac{1}{A^2-1} 
\left[ dx^a  dx^b ( u_a u_b + v_a  v_b - A (u_a v_b + v_a u_b))
\right]
\label{metrik2a},
\end{equation}
with
\[ u_a = (0,0,u_3,u_4) \quad v_a = (0,0,v_3,v_4), \]
where $-A = u_a v^a$ is the scalar product of the four velocities (
$|A|>1$). 
To ensure that both four velocities are geodesic they must be  gradients,
\begin{equation}
u_a = W_{,a} \quad v_a = V_{,a} .
\end{equation}
The scalars  $V$ and $W$ are arbitrary functions of $x^3$ and $x^4$. 

For explicit calculations two coordinate systems  were quite useful.
In the first one both velocities  have only one nontrivial covariant
component. With 
\begin{equation}
u_a =(0,0,0,-1) \quad v_a=(0,0,1,0) 
\label{metriku1}
\end{equation}
we get the line element \cite{Steph1}
\begin{equation}
ds^2 = Y^2 d \Omega^2 + \frac{1}{A^2-1} \left( dT^2 + dt^2 + 2 A dT dt
\right)
\label{metrik1}.
\end{equation}
In the coordinate chart (\ref{metrik1}) the energy momentum tensor is diagonal, 
so the field equations,  
$ G_{1 1 } = 0 $ and $G_{3 4} = 0 $, not involving the energy densities, are
given explicitly by a
 a coupled nonlinear system of PDEs
of second order:
\begin{eqnarray}
(A_t^2 + A_T^2) Y (- 2 A^2 -1) + A_t A_T Y A ( A^2 + 5)  & & \nonumber \\
+ (A_t Y_t + A_T Y_T) A ( A^2-1)+
(Y_t A_T + A_t Y_T) (1-A^2) & &  \nonumber \\ 
+ (A_{tt} + A_{TT}) Y A (A^2-1) + A_{tT} Y (1-A^4)   & & \nonumber \\
+ (Y_{tt} + Y_{TT}) (-(A^2-1)^2)+
Y_{tT} 2 A ( A^2-1)^2 & = & 0 
\label{PDE1}
\\
A^2 ( 2 Y_t Y_t + 2 Y Y_{tT})  & & \nonumber \\
 - A( Y_t^2+Y_T^2+ 2 Y Y_{tt} + 2 Y Y_{TT} +1)
+ 2 Y Y_{tT} &  = & 0. 
\label{PDE2}
\end{eqnarray}
The other non vanishing two components of the Einstein tensor  determine the energy densities:
\begin{equation}
G_{3 3} = \mu, \quad G_{4 4} = \epsilon. 
\label{density1}
\end{equation}

Another line element is of the Tolman--Bondi form, comoving to $u_a$
($W=-\bar{t}$):
\begin{equation}
ds^2 = Y^2 d \Omega^2 + \frac{V_r^2}{V_{\bar{t}}^2-1} dr^2 -d\bar{t}^2, \quad
u_a = (0,0,0,-1) \quad \mbox{and} \quad v_a = (0,0,V_r,V_{\bar{t}})
\label{metrik2}
\end{equation}
The line
elements (\ref{metrik1}) and (\ref{metrik2}) are related
by the transformation
\begin{equation}
T = V(r,t) \quad t = \bar{t}. 
\label{trafo1}
\end{equation}

In general both four velocities have got non zero shear and expansion.
But further restrictions can be obtained by using 
the Raychaudhuri equation (see \cite{Kra1}
and references therein)
for
each velocity and considering the field equations.
There are no solutions with positive energy densities 
where the expansion  is constant along the corresponding
vectors fields $u^a$ or $v^a$ respectively.
Additionally, after
a lengthy calculation 
 one can show that vanishing shear of {\it both}
$u_a$ and $v_a$  implies that the space time is flat \cite{Steph1,Wolf1}.

It is hopeless to find the general solution of the PDEs (\ref{PDE1}) and (\ref{PDE2}). One 
way to find explicit solutions is to look for additional symmetries in order to
reduce the number of variables. So the specific shape for  the free functions and the
radius $Y$ of the corresponding TBS is determined by the matching procedure. But
general properties and features of quite different models can be examined.
Before discussing this in detail in the sections 5 and 6,
we will consider the junction conditions between two-- and
one--component dust regions.

\section{Matching procedure}
\setcounter{equation}{0}
Here we discuss the matching between solutions with the energy
momentum tensors 
\[ T_{ab}^{(2)} = \mu \, v_a v_b + \epsilon \, u_a u_b \quad \mbox{and } \quad
T^{(1)}_{ab} = \rho u_a u_b  \]
across a time like hyper surface $\Sigma$. 
The matching procedure is based on the Darmois junction conditions
outlined in the work of Israel \cite{Isr1} and especially for dust
by Bonnor et.al. \cite{Bon2}. They require the continuity of the
first and second fundamental form.
$\Sigma$ has to be tangent to one of
the four velocities because there has to be no mass flux of the 
second dust component through the boundary surface.
As a consequence of the junction conditions 
 the sphere radius $Y$ and the free functions $M$
(\ref{mass1}) and
$f$ (\ref{para1}) of the Tolman--Bondi solution "produced" by the mass flux
through the boundary are continuous on $\Sigma$. 
Hence $M$ and $f$
for {\it one--component} dust can be calculated completely in
the {\it two--component} dust metric on $\Sigma$ because  
$M$ and $f$ are constant along the world lines of the corresponding dust component.
Vice versa a TBS for one--component dust gives the initial values for the
TCD on $\Sigma$.

The matching of two--component  dust 
to the vacuum is only  possible  if  both velocities
are parallel, i.e. in fact only for
one--component dust.
A continuous matching of the energy momentum tensor on $\Sigma$, that 
means
one of the two energy densities vanishes, is included
in the general scheme as a special case, too.

\section{Solutions  with an additional Killing vector}
\setcounter{equation}{0}
Special solutions with additional symmetries, i.e. 
Lie symmetries of the PDEs (\ref{PDE1})  and (\ref{PDE2}), lead to space times
which admit an additional Killing vector (or a homothetic vector, see section
6)  with only radial--time components. 
The Killing vector $\xi^a= ( 0,0,\xi^3,\xi^4)$
reflects a translational invariance in the
radial--time coordinates which do not appear explicitly
in the field equations. $\xi^a$ can be space like or
time like but not null because
$\xi^a \xi_a=0$ implies $\epsilon \mu < 0$.
Before discussing space like and time like Killing vectors 
separately in the next two subsections, we will examine common 
properties of the TCD admitting an extra Killing vector.

In the coordinate chart 
(\ref{metrik1}) the Killing vector $\xi^a$ is
\begin{equation}
\xi^a = \left(0,0,1,-\frac{1}{k}\right)   
\quad \xi^a \xi_a = \frac{k^2 +1 - 2 k A}{k^2 (A^2-1)},
\label{killan1}
\end{equation}
where the  metric coefficients depend only on the reduction variable $z=T+kt$
with a constant $k \neq 0$:
\begin{eqnarray}
A & = & A(T + k t) \nonumber \\
Y  & = & Y(T + k t).
\label{ansatz1}
\end{eqnarray}
The field equations $G_{11} =0$ (\ref{PDE1}) and
$G_{34}=0$ (\ref{PDE2})  now read
\begin{eqnarray}
A^2 2 k ( Y Y'' + (Y')^2)  \nonumber \\
+ A ( -1 - (Y')^2 (1 + k^2) - 2 Y Y'' (1 + k^2))
+ 2 k Y Y'' & = & 0
\label{feld1}\\
Y (A')^2 ( (1+k^2)(-1 - 2 A^2) + k A ( 5 + A^2)) \nonumber \\ +
A' Y' (A^2-1)(-2k + A + k A ) & & \nonumber \\
+ Y A'' ( ( A^2-1) (A + k^2 A ) - k (A^4-1) ) \nonumber \\ +
Y''((A^2-1)^2(-1 - k^2
+ 2k A)) & = & 0.
\label{feld2}
\end{eqnarray}
They have reduced to  an autonomous system of ODEs where the prime denotes
differentiation w.r.t. $z$.
The energy densities obtained from (\ref{density1}) become 
\begin{eqnarray}
\mu & = & \frac{E(z)}{(-1 + k A) Y^2 (A^2-1) }
\label{energie1} \\
\epsilon & = & \frac{k E(z) }{(-k + A) Y^2 (A^2-1)}
\label{energie2},
\end{eqnarray}
when eliminating the second derivatives with the aid
of the field equations (\ref{feld1}) and (\ref{feld2}).
Because of the common factor $E(z)$ in  the  two energy densities
(\ref{energie1}) and (\ref{energie2}) it is not possible  that only one
vanishes, i.e. the Tolman--Bondi solution cannot admit such a Killing vector
except in the vacuum case. As a consequence the geometry and topology
of the corresponding TBS differs from the two--dust--component.

 When the Killing vector is orthogonal to one of the four velocities, i.e. $k= 0$ or $k=\infty$,
one gets one--component dust which
belongs to the class of
Kan\-tows\-ki--Sachs
because the Killing vector is space like \cite{Kra1}.

In the coordinate system (\ref{metrik2}),
comoving to $u^a$,   
the line element
can be transformed in a form similar to (\ref{ansatz1}):
\begin{equation}
ds^2 = F_1^2(\tilde{z}) d\Omega^2 + \frac{F_2^2(\tilde{z})}{
    (\tilde{k} F_2(\tilde{z}) +c_1)^2 -1} dr^2 -dt^2 \quad
    \tilde{z} = r + \tilde{k} \, t
\label{metrik3}
\end{equation}
with
\begin{equation}
u_a=(0,0,0,1), \quad v_a = (0,0,F_2(\tilde{z}),
    \tilde{k} F_2(\tilde{z}) + c_1) \quad \mbox{and} \quad
 \xi^a = (0,0,1,-\frac{1}{\tilde{k}}).
 \end{equation}

Because of the existence of an additional Killing vector the free functions $M$ and $f$ and the 
sphere radius $Y$ of the corresponding solution for one dust component are determined completely
by the Darmois junction conditions. So one knows the properties of the corresponding TBS by
calculating the metric coefficients of the TCD.

\subsection{Solution with a space like Killing vector}

Solutions of two--component dust with a space like Killing vector,
i.e a group $G_4$ acting on a $S_3$ (notation see \cite{Kra1}),
are generalizations of the Kantowski--Sachs class for one--component dust.
Here an Euclidean "center" of symmetry is lacking. A radial 
coordinate can extend from $-\infty$ to $\infty$. Such models
for dust and perfect fluid (where $Y$ depends only on the comoving
time) are also called T--models; see Ruban \cite{Rub1} and references
therein for a deeper discussion. 

Possible singularities in the energy
densities might occur for $Y=0$, which 
could be interpreted as
a "big bang" or  gravitational collapse "big crunch".
As an approximation of first order for a singularity at $z=0$ for the
reduction variable one gets 
with the aid of the field equations 
\[ |A| \approx  1+\frac{9}{8 c k^2} z^{2/3} ,\quad Y \approx c z^{2/3} \quad
\mbox{and} \quad \epsilon \sim \frac{1}{z^{4/3}}, \quad  \mu 
\sim \frac{1}{z^{4/3}}. \]
So $|A|=1$ and $Y=0$ occur simultaneously, i.e. in the neighbourhood
of the big bang singularity  both four velocities are nearly parallel.
Here the asymptotic is the same as for one dust component with
Kantowski--Sachs geometry.
The expansion is stopped  when the TCD reaches its own
Schwarzschild sphere because  the derivative $Y'$ vanishes for $Y=M$ when
substituting the ansatz (\ref{ansatz1}) in the formula (\ref{mass1}) for the mass function $M$. Then
the re--collapse begins and ends in the final big crunch singularity
(see Figure 1). The T--models for dust and perfect fluid
(especially the "uniform T--models") discussed in \cite{Rub1}
are completely analogous; so that solution for itself can be
interpreted as a generalization of the
uniform T-models for two--component--dust and represents a closed
universe.
To get a concrete solution for the TCD and the corresponding TBS the
ODEs (\ref{feld1}) and (\ref{feld2}) have to be solved numerically.
The initial values determine the
maximal expansion, 
the relative velocity and the values of the energy densities
for each dust component in a nonlinear way.

Because of the range of $Y$
this model can be interpreted as central core
consisting
of two--component dust related to one--component dust outside due to
a mass flux. 
The mass function $M$ is not a monotonic function (see Figure 2). The decreasing part
indicating a mass loss for the TCD region is separated from the
increasing part (ingoing mass flux) by the "turning point" 
$Y=M$.
 The ingoing and
outgoing mass flux is not balanced because
there is a sign change for the "energy function" $f$
(see Figure 2).
So a part of the exterior
TBS is connected to an open model while otherwise
$f$ lead always to a closed model ($f<0$).
The properties for $f$ and $M$ can be compared to the
 "neck--and--two--sheets"
topology  with the same qualitative features for the TBS
discussed by Hellaby \cite{Hell2}.

\subsection{Solution with a time like Killing vector}
For a time like Killing vector the metric
is static. There exists no corresponding non--vacuum solution  for
one--component dust. 
The type of shell--crossing singularities 
 can occur
when $\xi^a$ 
becomes parallel to $u^a$ or $v^a$.
Here the corresponding energy density is diverging
but the metric itself and the other energy density
remain finite for a non--comoving coordinate system. 
Because the condition $Y\ge M$ is fulfilled for time like Killing vectors
the sphere radius $Y$ has its maximal and minimal values when
the shell--crossing singularities occur.
To get a quantitative solution the ODEs (\ref{feld1}) and (\ref{feld2}) have to be solved numerically.
Numerical solutions show that the other energy density also
reaches its extremal values at the shell--crossing singularities (see figure 3),
e.g.
\[ Y_1 < Y < Y_2 \quad \mbox{and} \quad \mu_1 < \mu < \mu_2. \]
The constants $Y_1$, $Y_2$, $\mu_1$ and $\mu_2$ are determined by the
initial values.

Because of the finite values
of $Y$
this model can be interpreted as a thick shell consisting of two dust components.
The shell is "produced"
by two shells of one-component dust where the thickness of the shells
 is restricted by shell--crossing
singularities.
First the shells are separated then they  crash together and
penetrate through each other. So the
interior of the "mixing zone" is static but the boundaries are varying 
in time. 
The two inner boundaries of the shells meet each other
when the mass function $M$ (figure 4) vanishes, let us say for $Y= \bar{Y}$,
because  the
junction conditions for the flat space has to be fulfilled for a moment.
Before this moment the normal vector of the surface is orthogonal to e.g.
$u^a$ then orthogonal to $v^a$. 
Because the mass function is
negative  a TCD shell cannot be matched for
$Y_1 < Y < \bar{Y}$. The shells
have to re--separate because 
the dust components never become comoving.
The energy function $f$ for $v_a$ (Figure 4)  changes from positive to negative values.
An ever expanding sphere is slowed down due to the shell crossing.
That can lead to a bound state and end in an gravitational collapse.
The other dust shell belongs to a closed model.

\section{Solutions with an additional homothetic vector}
\setcounter{equation}{0}
A homothetic vector $\eta^a$ 
 represents a self--similar behaviour in the
radial--time coordinates of the metric.
Eardley \cite {Eard1} examined self--similar space times in general.
Cahill and Taub \cite{Cah1} considered spherically symmetric 
self--similar space times for a perfect fluid, especially
they investigated self--similar dust solutions.
In recent years self--similar space times were extensively studied, 
for dust see e.g.
 \cite{Pon1}. The self--similar dust solution is included 
 here as a particular solution.
The main
difference to the solutions of the previous section is that
$\eta^a \eta_a$ can change sign
for the {\it same} solution, i.e. $\eta^a$ can be space like or 
time like in different regions.

In the coordinate system (\ref{metrik1})
the explicit expression for the homothetic vector is
\begin{equation}
\eta^a = ( 0,0,T,t),  \quad 
\eta_a \eta^a = \frac{t^2}{G^2-1} \left( 1+z^2 + 2 z G \right)
\label{homo1}
\end{equation}
with the line element
\begin{equation}
ds^2 = t^2 F^2(z) d\Omega^2 + \frac{1}{G^2(z)-1} \left(
dT^2 + dt^2 + 2 G(z) dT dt \right) \quad  z= \frac{T}{t}
\label{metrik5a}.
\end{equation}
The field equations (\ref{PDE1}) and (\ref{PDE2}) are thus reduced to 
a non--autonomous system of ODEs
\begin{eqnarray}
G G' F  (z +G) ( G^2-1)  \nonumber \\+ F (G')^2 (-1 -z^2 - 5 z G  - 2 G^2 - 2 z^2
 G^2
  - z G^3) & & \nonumber \\
  + G' F' ( 2 z + G + z^2 G) (G^2-1) \nonumber \\
  + G'' F ( (G +z^2 G )(G^2-1) + z (G^4-1))
  \nonumber \\
   + F''  (G^2-1)^2(-1 -z^2 - 2 z G)  & = & 0 
\label{homoode1} \\
G + G F^2 - 2 z G F F' - 2 G^2 F F'  & & \nonumber \\
+ G (F')^2 + z^2 G (F')^2 +
2 z G^2 (F')^2 & & \nonumber \\
+ 2 z F F'' + 2 G F F'' + 2 z^2 G F F'' + 2 z G^2 F F'' & = & 0
\label{homoode2}
\end{eqnarray}
which cannot be solved analytically. 
The energy densities can be expressed
in the form
\begin{eqnarray}
\epsilon & = & \frac{1}{t^2 (G^2(z)-1) F^2(z)} \, \epsilon_1(z) \\
\mu & = & \frac{ 1}{t^2 (G^2(z)-1) F^2(z)} \, \mu_1(z) 
\end{eqnarray}
for some functions $\mu_1$ and $\epsilon_1$ depending on the reduction
variable $z$ alone. 
The energy densities do not vanish except for the particular
TBS. 
In the comoving
coordinate system (\ref{metrik2})
the line element can be written in a form analogous to (\ref{metrik5a}).

The qualitative behaviour for the solutions
of the field equations (\ref{homoode1}) and (\ref{homoode2}) can be compared
to the solutions discussed in the previous section. 
A big bang singularity with space like $\eta^a$ as well as a shell--crossing
singularity with  time like $\eta^a$  can occur both.
The coordinate dependence of the singularities show the same asymptotic.
But in contrast to the space like Killing vector with Kantowski--Sachs
geometry
a sphere
can expand through  its own Schwarzschild sphere for the closed model with big bang
 and big crunch singularity.
So the mass
function $M$ is a strict monotonic function  indicating that
there always is a mass loss through the boundary generating
the TBS  outside.
The phenomena with two shells
of dust penetrating each other and re--separating is possible, too.
Here the homothetic vector is time like for the TCD region. 

In addition to the already known models  space times admitting
a homothetic vector contain another type of solutions,
presenting 
a compact ball of one--component dust
inside surrounded by a shell of dust with finite size outside.
The
dust shell is crashing on the dust ball and  ends  in
a gravitational collapse. Here the space time is restricted by 
a big bang (big crunch) and a shell--crossing singularity
where the maximal radius $Y$ is reached (see Figure 5).
The mass function $M$  of the corresponding one--component dust is a positive monotonic 
function, i.e mass flux in one direction.
The energy function $f$ 
leads
to an elliptic (closed) TBS w.r.t. to both $u^a$ or $v^a$.

\section{Conclusions}
As a generalization of the well--known Tolman--Bondi solution for 
spherically symmetric dust the model of two--component dust in radial motion
was introduced.
The corresponding field equations were formulated in an invariant way.
The connection of the two--component dust to the Tolman--Bondi solution
was discussed and the matching procedure outlined.
This                                                        
model can describe  different physical models and topologies.
Two--component dust with a time like Killing vector can describe the crossing of thick 
shells of dust where the thickness of the shells is restricted by shell--crossing singularities.
A space like Killing vector leads to a central core of two dust components
 with Kantowski--Sachs
geometry and ingoing and outgoing mass flux. The topology of the corresponding Tolman--Bondi
solution outside can be  compared with the "neck--and--two--sheets" topology
discussed by Hellaby. In the case with the homothetic vector the mass flux can go only
in one direction. Additionally big bang and shell--crossing singularities can occur 
simultaneously leading to the model with a ball of dust surrounded by a dust shell.  

\section*{Acknowledgments}
The author would like to thank Prof. H. Stephani for the introduction
into this interesting topic and for fruitful
discussions.

\newpage
\pagestyle{empty}
\unitlength=1cm
\begin{figure}
\begin{picture}(10,20)
\put(-1.5,-1){
\psfig{figure=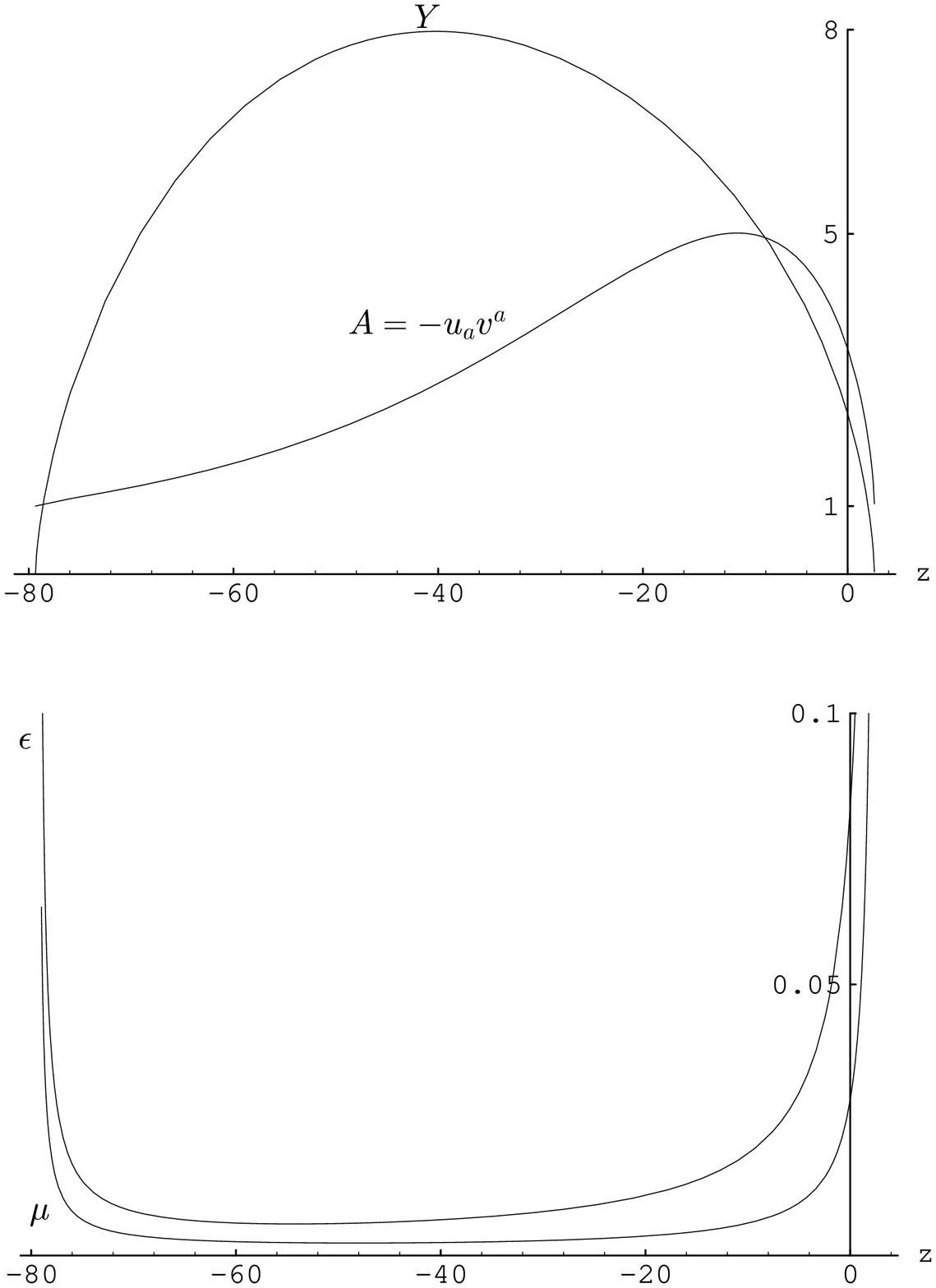,height=25cm,angle=0}}
\end{picture}
\caption{ 
Metric coefficients and energy densities for a
solution of two--component  dust with a space like Killing vector
$\xi^a = (0,0,1,-1/2)$
 and  the initial
 values
  $Y(2)=1$, $Y'(2)=-1$, $A(2) = 2$ and $A'(2)=
  -1$. In the neighborhood of the big bang (big crunch) singularity  
  both dust components
  are nearly comoving indicated by $A=1$.
}
\end{figure}
\newpage
\unitlength=1cm
\begin{figure}
\begin{picture}(10,20)
\put(-1.5,-1){
\psfig{figure=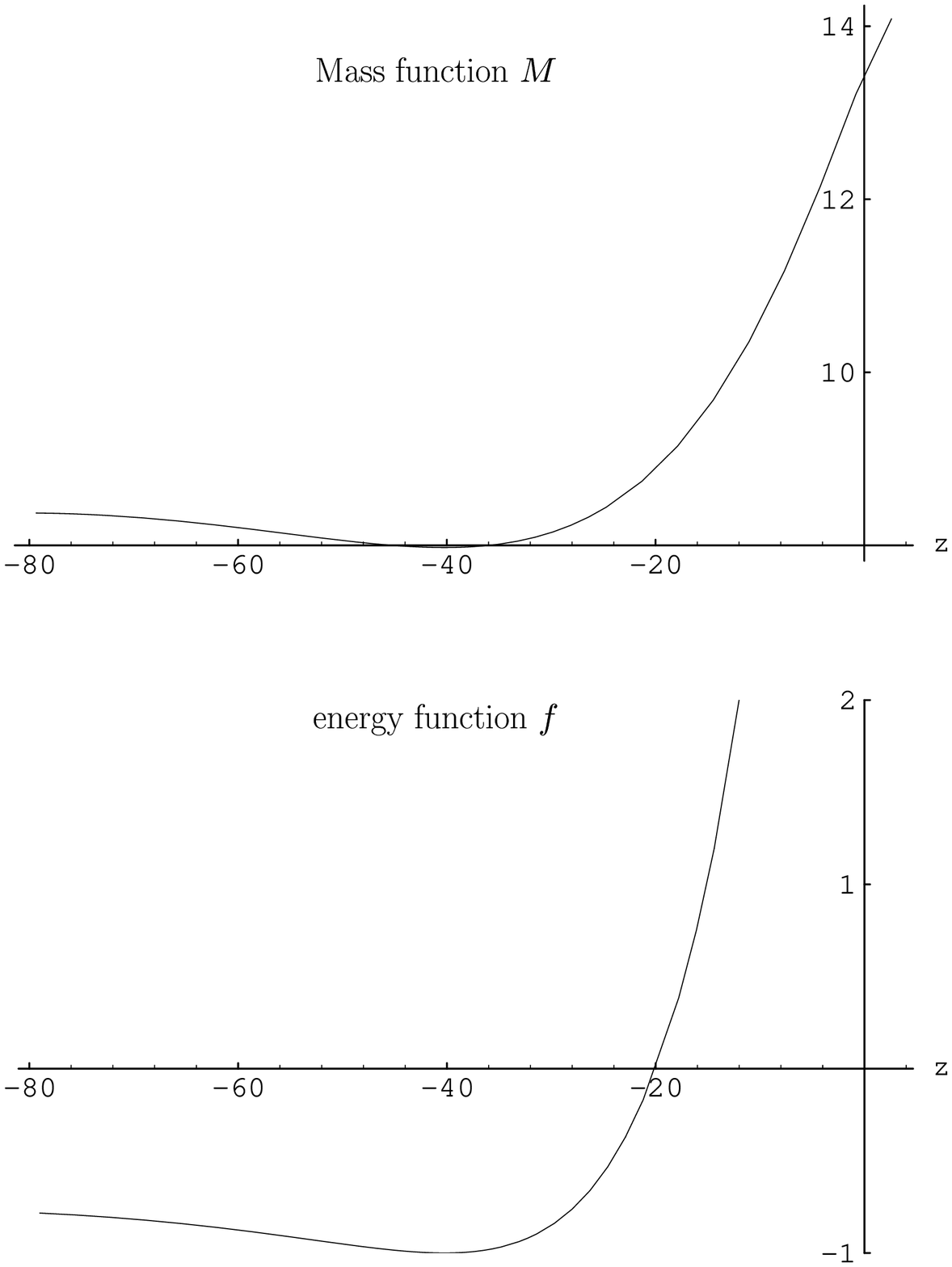,height=25cm,angle=0}}
\end{picture}
\caption{
The mass function $M$ of the exterior one--component dust
is not monotonic because of ingoing and outgoing mass flux.
 The mass flux is not balanced because
 the sign change of the energy function $f$ indicates that a part 
 belongs to an open
 model.The energy function $f$ belonging to the other four velocity 
 is only multi
 plied by the factor $k$.}
\end{figure}
\newpage
\unitlength=1cm
\begin{figure}
\begin{picture}(10,20)
\put(-1.5,-1){
\psfig{figure=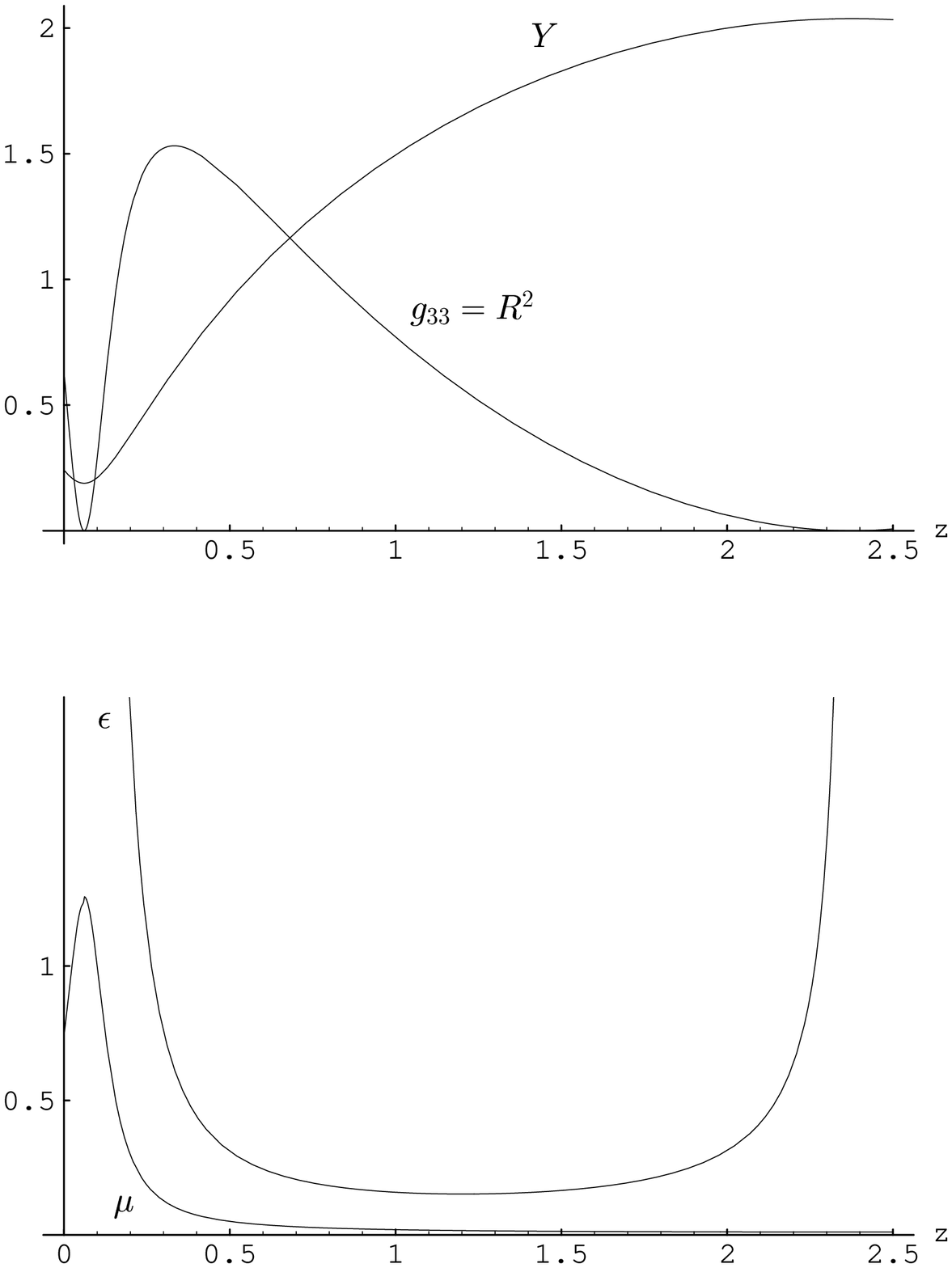,height=25cm,angle=0}}
\end{picture}
\caption{
Metric coefficients  and energy densities
 for two--component dust with a time like Killing vector
 $\xi^a = (0,0,1,-2)$ in the coordinate chart (\ref{metrik3})
  comoving to $u^a$:
   with the parameter $c_1=-2$
   and the initial values $F_1(2) =2$, $F_1'(2)=1/5$,
   $F_2(2)= -1/2$ and $F_2'(2)=3/2$. In this coordinate chart
   the shell--crossing
   singularity with diverging $\epsilon$ occurs at $R=0$
   where $Y$ and the other energy density
   $\mu$ has got its extremal valuesa.}
\end{figure}
\newpage
\unitlength=1cm
\begin{figure}
\begin{picture}(10,20)
\put(-1.5,-1){
\psfig{figure=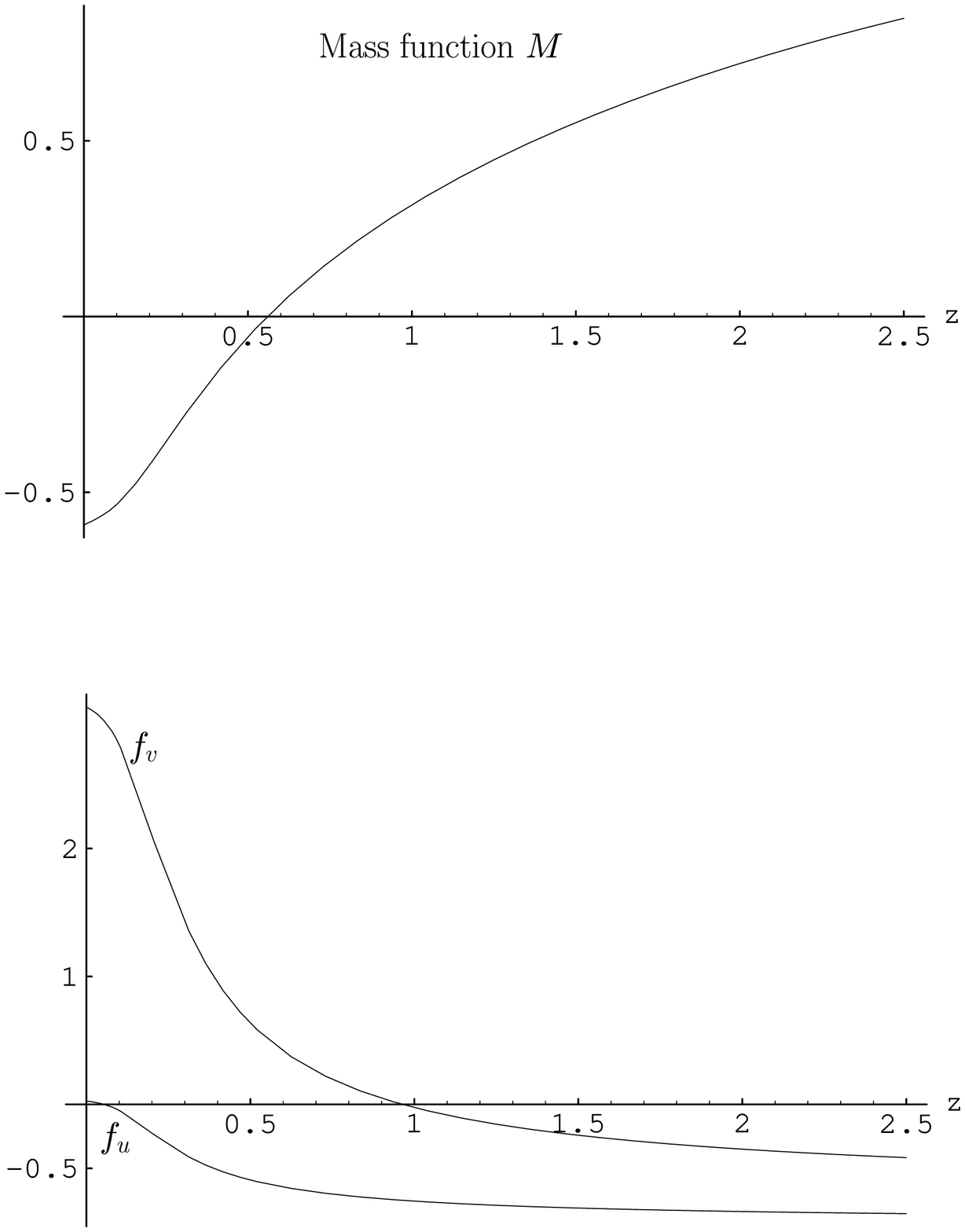,height=25cm,angle=0}}
\end{picture}
\caption{
Free functions for the corresponding dust component:
$M=0$ belongs to the moment   when
the inner spheres of the shells meet each other.
While $f_u$, denoting the energy function w.r.t. $u_a$, belongs to
a closed model $f_v$ changes sign, i.e.
an ever expanding  dust shell inside
is slowed down  due to the shell crossing and can re--collapse
afterwards.
}
\end{figure}
\newpage
\unitlength=1cm
\begin{figure}
\begin{picture}(10,20)
\put(-1.5,-1){
\psfig{figure=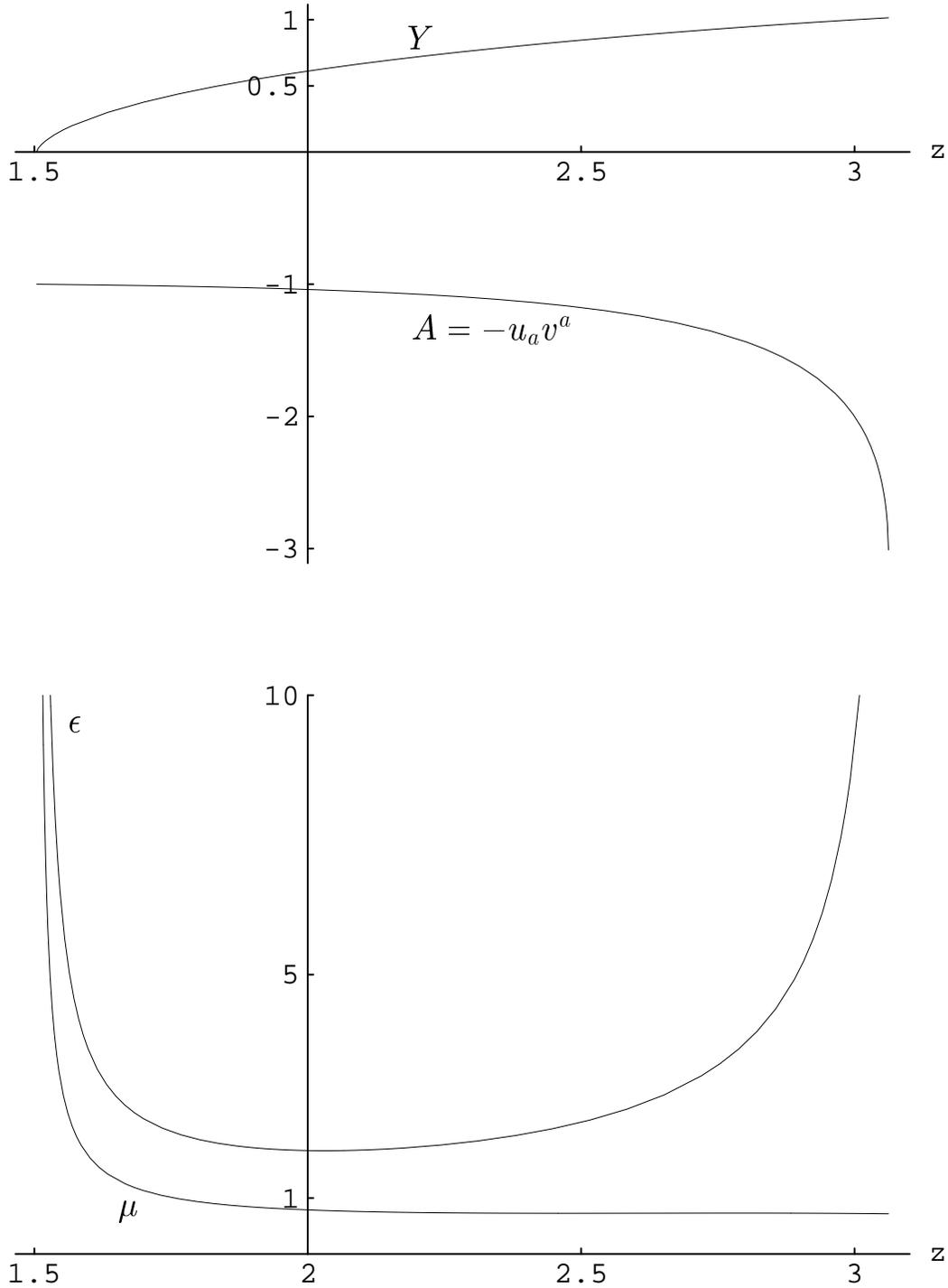,height=25cm,angle=0}}
\end{picture}
\caption{
Metric coefficients and energy densities
with the initial
values
$F(3) = 1$, $ F'(3) = 1/4$, $ G(3) = -2$ and $G'(3) = -6$: The
scaling factor $t$ has been set to unity.
A big bang (big crunch) singularity occurs at $Y=0$.
The energy density $\epsilon$ diverges because of a shell--crossing
singularity when
the homothetic
vector becomes parallel to the four velocity $u^a$.
}
\end{figure}

\end{document}